# Geochemistry of U and Th and its Influence on the Origin and Evolution of the Earth's Crust and the Biological Evolution

Xuezhao Bao[a1] and Ali Zhang

[a] Department of Earth Sciences, the University of Western Ontario, 34-534, Platt's Lane, London, Canada, N6G 3A8

**Abstract:**

We have investigated the migration behaviors of uranium (U) and thorium (Th) in Earth and other terrestrial planets. Theoretical models of U and Th migration have been proposed. These models suggest that the unique features of Earth are closely connected with its unique U and Th migration models and distribution patterns. In the Earth, U and Th can combine with oxidative volatile components and water, migrate up to the asthenosphere position to form an enrichment zone (EZ) of U and Th first, and then migrate up further to the crusts through magmatism and metamorphism. We emphasize that the formation of an EZ of U, Th and other heat-producing elements is a prerequisite for the formation of a plate tectonic system. The heat-producing elements, currently mainly U and Th, in the EZ are also the energy sources that drive the formation and evolution of Earth's crust and create special granitic continental crusts. In other terrestrial planets, including Mercury, Venus, and Mars, an EZ can not be formed because of a lack of oxidative volatile components and water. For this reason, a plate tectonic system can not been developed in these planets. We also emphasize the influence of U and Th in EZ on the development and evolution of life on Earth. We propose that since the Earth and planets were born in a united solar system, there should be some common mechanisms to create the similarities and differences between them. We have tried to develop an integrated view to explain some problems in the Earth's tectonics and evolution, bio-evolution, and planetary dynamics through U and Th geochemistry. We believe that a comprehensive exploration on energy sources and their evolution is a good way to build bridges between different disciplines of science in order to better understand the Earth and planets.

## 1. INTRODUCTION

U, Th and K are the main heat-producing elements in planetary materials. The heat released from them is the major energy source to drive the evolution of Earth and planets [1], and also the major power to drive the origin and evolution of the continental crust on Earth [2]. In the Earth, the heat released from $U^{235}$ is $7 \times 10^{18}$ cal/year, from $U^{238}$ 171.8 x $10^{18}$ cal/year, from Th 173 x $10^{18}$ cal/year, and from K(potassium) only 73.83 x $10^{18}$ cal/year [3]. Therefore, the content, existing status and migration patterns of U and Th have an important impact on the evolution of planets. At the same time, U and Th are radioactive. The concentration variation of U in sea water might have caused the extinction of some species [4]. Thus, it is very important to investigate the migration behaviors of U and Th to understand planetary evolution, the formation and evolution of Earth's crust, and bio-evolution.

---

[1] Corresponding author. E-mail address: xuezhaobao@hotmail.com (Xuezhao Bao)

## 2. THE PHYSICAL AND CHEMICAL PROPERTIES OF U AND Th AND THEIR INFLUENCE ON THEIR MIGRATION

As mentioned above, U and Th are the major energy sources that drive the evolution of Earth and planets. Thus it is very important to investigate their existing status and migration behaviors. From table 1, we can see: (1) metal U and Th have high melting points and high densities. Therefore under ultra-reducing conditions, they are stable and reluctant to migrate. (2) Their low valance oxides and compounds also have high densities and even higher melting points. For instance, $UO_2$ and $ThO_2$ have melting points of 2800 $^oC$ and 3220 $^oC$ respectively. Therefore, under reducing conditions, U and Th are also stable and difficult to migrate. However, their high valence compounds and complexes have low melting points, small densities, and are water-soluble. Therefore, under oxidative conditions, they are prone to migration. The following is a description of ways in which U and Th can migrate: (1) they can be oxidized by elements such as F, Cl etc. in the halogen family with a strong oxidative ability and form compounds and complexes with low melting points and volatility [6]. (2) They can be oxidized by $O_2$, $H_2O$, $N_2$ etc, to form $UO_3$ [8, 9] and other compounds and complexes with low melting points and volatility. Under oxidative condition, U can form water soluble $UO_2^{2+}$ ionic group. (3) They can combine with the $CO_3^{3-}$, $SO_4^{2-}$, $NO_3^-$ acid root ionic groups to form complexes and migrate [5,6,10,11]. These acid root ionic groups need water and oxygen to form: $C + O_2 \rightarrow CO_2$, $CO_2 + H_2O \rightarrow 2H^- + CO_3^{2-}$; $2S + 3O_2 \rightarrow SO_3$, $SO_3 + H_2O \rightarrow 2H^- + SO_4^{2-}$... (4) Under oxidative conditions, U and Th will become high valence ions, which make them easier to combine with other elements or ionic groups to form complexes. Moreover, the higher the valences of U and Th, the more anions or anion groups they can combine with to form complexes [12]. Consequently, the complexes of U and Th will become smaller in density, and easier to migrate up. The fact that an increase of isolated $O^{2-}$ from ultra-mafic, mafic, intermediate to felsic magmas [13] is accompanied by an increase in U and Th content indicates the importance of $O^{2-}$ in the migration and enrichment of U and Th. Thus, how U and Th migrate depends on whether or not oxygen, oxidative volatiles and $H_2O$, or their ions exist. As pointed out by Liu [6], the migration of U and Th depends on how much volatile components the rocks have. Also, only volatile elements and their complex anions with a small density can lower the density of U and Th compounds and complexes and allow U and Th to move up and enrich in the Earth's crust.

In addition, U and Th have big cation radii, small Electronegativities and ionization potentials ($U^{4+}$: 9.7 nm, 1.4, 6.08 eV respectively; $Th^{4+}$: 11.4 nm, 1.0, 6.95 Ev respectively). Therefore, their ions are similar to $K^+$, $Na^+$ and rare earth elements (REE) in these aspects [5,6], and are prone to form compounds with weak chemical bonds and migrated easily. Consequently, U and Th are usually enriched in rocks that are also rich in $K^+$, $Na^+$ and REE ions [11, 14].

**Table 1 The melting points (°C) and densities (g/cm$^3$) of U, Th, their compounds and complexes and core related elements***

| Element | Density | Melting point | Compound | Density | Melting point | Compound | Density | Melting point |
|---|---|---|---|---|---|---|---|---|
| Fe | 7.86 | 1535 | $ThO_2$ | 10.0 | 3220 | $UO_2$ | 10.95 | 2800 |
| Ni | 8.90 | 1453 | ThN | 10.6 | 2500 | UCl | 4.86 | 567 |
| U | 19.05 | 1132 | ThS | | 1905 | UN | 14.32 | 2800 |
| Th | 17.70 | 1700 | $ThCl_4$ | | 600 | $UF_6$ | 4.68 | ~0 |
| $K_4ThOX_4\cdot 4H_2O$ soluble in water | | | $Th(NO_3)_4\cdot 5H_2O$ soluble in water | | | $UO_3$ | 7.29 | soluble in water |

**\*Data come from refs 5-9.**

## 3. THE MIGRATION MODELS OF U AND Th IN THE EARTH'S UPPER PART AND THEIR INFLUENCE ON THE CRUSTAL EVOLUTION

Oceanic water can be carried down to the lower mantle [15], and the rocks with volatile compounds in the oceanic crust might be transferred to the core-mantle boundary (CMB) by subduction in a plate tectonic system [15, 16]. Moreover, the amount of water in the mantle may be three times more than that of the oceans [15]. Therefore, in the range of the crust and mantle (upper mantle), the rocks contain water, volatile elements or their ions. Consequently, U and Th can exist as compounds and complexes, and migrate up easily as they are heated.

3.1 The geochemistry behaviors of U and Th during magma crystallization

Magmatic rocks are ones of the major constituents of the Earth's crust. From ultra-mafic (formed at high temperature (T)), mafic (high to medium T), intermediate (medium T) and felsic (low T, including alkali rocks) rocks, the concentration of U and Th generally increases [5, 6, 11, 14]. The chemical zonation of zircons from Magmatic rocks clearly indicates that the content of U and Th in the remnant magma gradually increases during crystallization and differentiation [17-21]. Similar zonations can also be found in zircons crystallized in a mafic magma with a low U, Th content [21]. Magma crystallization and differentiation is a process of decreasing temperature. Thus U and Th have difficulty entering mafic mineral crystals in an early high temperature stage, but are enriched in the rocks with more felsic compositions crystallized in a late low temperature stage. The data on the relationship between crystal shapes (forms) and formation temperatures of zircons from granites also indicate that U and Th are enriched in the zircons crystallized in a late low temperature stage [22] as indicated in fig.1.

3.2 The U, Th geochemistry of meteorites

The analyses of chondrites show that the concentration of U is lowest in the big chondrules, moderate in the small chondrules, and highest in the matrix around these chondrules [23]. Namely from the centre of chondrules to the matrix at their edges, the U concentration increases, which is similar to the trend of U+Th concentration increasing from the core to rim of a magmatic zircon crystal [18]. Magmatic zircons crystallize during a process of decreasing magma temperature (condensation). Therefore, meteorites also form in a process of decreasing temperature.

In addition, U and Th concentrations are low in the iron meteorites, medium in the iron-stone meteorites and relatively high in the stone meteorites [10]. According to meteoritics, iron meteorites have usually formed in areas with small heliocentric distances (the distance from the Sun), and in the mid-plane of the Sun's protoplanetary disk. In these areas, temperatures were usually high. On the other hand, stone meteorites have formed in areas with large heliocentric distances, and in areas far away from the mid-plane of the Sun's protoplanetary disk. In these areas, temperatures were usually low [15, 24]. This indicates that U and Th concentrations are low in high temperature areas (iron meteorites), and relatively high in the low temperature areas(stone meteorites) in Sun's protoplanetary disk, which is consistent with the U, Th variation trend from the high temperature ultra mafic rocks to low-temperature felsic rocks mentioned above. From H, L to LL-type meteorites correspondingly, the relative locations they are formed at range from small heliocentric distances to large heliocentric distances, the U and Th concentrations increase [23]. This also clearly indicates that the migration of U and Th is from high to low temperature areas in the Sun's protoplanetary disk.

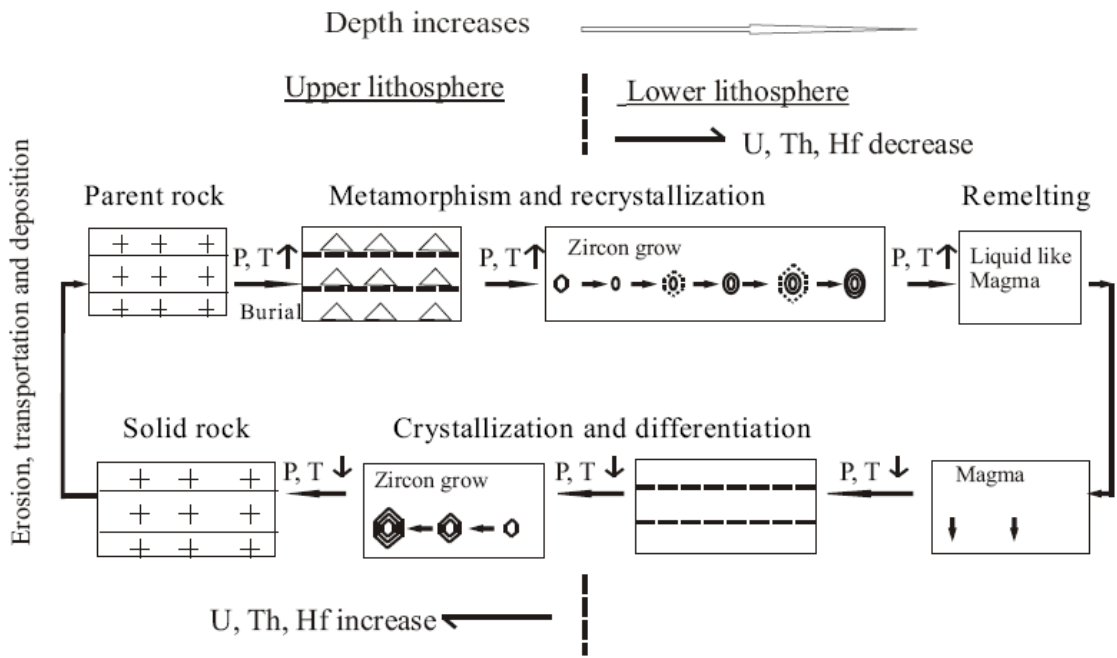

**Fig.1 The migration of U, Th and Hf in magmatism and metamorphism.**
**Upper part: metamorphism: Zircon exhibits rounded growth zoning; U and Th decrease from core to rim.**
**Lower part: magmatism: Zircon exhibits euhedral growth zoning. U and Th increase from core to rim or from inner to outer parts within each small composition zone**

3.3 The geochemistry behaviors of U and Th in metamorphism

Metamorphic rocks are also ones of the major constituents of the Earth's crust. We have found that the (U + Th) concentration from core to rim of metamorphic zircon crystals decreases generally [17,18], which shows that with the metamorphic temperature increasing, the (U+Th) concentration of the parent rocks that zircons crystallized from gradually decreases. In high metamorphic temperature stage, the zircons with very low concentrations of U and Th can be produced [21]. This U and Th variation trend is similar to that of meteorites: when they were heated, the U concentration gradually decreased with an increase of temperature. Namely, rising temperature leads U to migrate from meteorites [22]. Therefore, metamorphically heating rock and manually heating meteorites will both lead to the migration of U and Th from the parent rocks or meteorites. This indicates that U and Th are active, and their ions are prone to migrate when being heated.

During the subduction of lithosphere plates, a metamorphic process with an increase in temperature and pressure will occur in the subducting materials. The field observations and corresponding experimental studies of Tatsumi et al.[25] show that with the subduction, the temperature and pressure of the subducting plates (downgoing plates) increases, and many ions become active and migrate from the subducting plates. Their migration rates have a direct positive relationship with the radius of ions. Namely, the ions with a larger radius migrate up more easily during metamorphism with a rise in temperature and pressure [25]. The ionic radii of U and Th are much larger than those of mafic ions, such as Fe, Ni etc. [5,6]. Therefore, they are prone to migrate from the parent rock with K, Na etc that also have large ionic radii. Consequently, U and Th can not return to the mantle in a significant amount when the oceanic plates subduct to the mantle, but migrate up to continental crusts with metamorphic hot liquid or magma produced during the subduction. This is also a very important process for the enrichment of U and Th in continental crusts as illustrated in fig.1.

Summary: U and Th migrate easily from high temperature to low temperature phases when there are oxidative volatile components, water or their ions. Based on the current knowledge, the temperature increases gradually from the surface to the deep interior of our planet. Therefore, in the range of the upper mantle, U and Th will gradually migrate to the surface. The top layers of the Earth consist mainly of granitic rocks. The fact that the U and Th contents of granitic rocks are highest among rocks in the Earth strongly supports the migration trend of U and Th described above.

3.4 The evolution of U and Th in crustal rocks

From fig. 1 and the discussion above, there is a tendency for U and Th to migrate from Earth's interior to the surface. This kind of migration is completed by magmatism and metamorphism. In other words, the U and Th contents in the crust increase with geologic time.

Table 2 lists the U and Th contents of accessory minerals from Phanerozoic granitoids and Precambrian granitoids. It shows that the U and Th contents increase with geologic time [26]. Other studies also support the ideas that the same type of rocks with younger ages are richer in U and Th than those with older ages [5,6]. Therefore, the major evolution trend is that U and Th gradually migrate up from the deep interior, including the core, to the crust.

3.5 The relation between U, Th migration model in Earth's upper part and the evolution of Earth's crust

The discussion above clearly indicates that the crusts, especially continental crusts, are rich in U and Th and their U and Th contents increase with geologic time. This

**Table 2 The average contents of U, Th and REE (Rare Earth Elements) in Precambrian (I) and Phanerozoic (II) granitoids (gram/ton) ***

| Elements | Allanite | Apatite | Zircon | Titanite/sphene | Barringerite |
|---|---|---|---|---|---|
| REE I | 198900.0 | 3514.3 | 3050.0 | 14457.1 | N/A |
| REE II | 210252.8 | 6629.8 | 4895.4 | 17694.1 | N/A |
| U I | 453.0 | 32.5 | 900.0 | 56.6 | N/A |
| U II | 515.2 | 92.8 | 1150.2 | 208.1 | 15 |
| Th I | 7525.0 | 52.8 | 400.0 | 1 | N/A |
| Th II | 7224.4 | 171.8 | 913.3 | 471.2 | 71.5 |

*** data come from ref 26.**

increase of U and Th is usually accompanied by an increase in K, Na and rare earth (REE) etc., which are incompatible elements with big ion radii. This is also determined by their similarity in crystal chemical properties.

However, the kimberlites coming from 400-650 km in the mantle [27], ultra-potassium volcanic rocks that are similar to kimberlites, and alkali basalts, are also rich in these elements [11, 14]. Furthermore, the REE distribution pattern of kimberlites (including some alkali basalts) indicate that they are richer in light REE, and their La/Yb ratios are greater than or are equal to that of granites[11, 14]. This signifies that these elements in the source magmas of kimberlites may have undergone a complete differentiation, which is very different from normal mantle rocks, such as peridotite, olivine-pyroxene rocks (pyrolite) [14]. This means that before they reach the 400-650 km transition zone, the source magmas of these rocks have experienced relatively good differentiation. Therefore, this transition zone was considered to be an original crust [28, 29].

3.6 Possible models

It is widely accepted that Earth formed through a process of accretion from planetesimals, according to planetesimal theory [28, 30]. The original Earth is considered to have accreted from several planetesimals (or planetary embryos) of radius > 1000 km, dozens of planetesimals of radius > 100 km, and hundreds of planetesimals of radius >10 km [30] as illustrated in fig.2.

Among the planetesimals, those with a radius of > 100 km might have differentiated into chemical layers from core, mantle to crust, or from core to mantle as showed in fig.2 A. As discussed above, the nebula materials in the Sun's protoplanetary disk might have differentiated before forming planetesimals. During the formation of planetesimals, iron meteorite components with large densities and low U, Th contents [10] formed their cores, and stone meteorite components with small densities and relatively high U and Th contents [10] formed their crusts[15, 24]. At the same time, the impact energy and decay energy released from the radioactive elements inside planetesimals also caused them to differentiate into core, mantle and crust. As illustrated in fig.2, after these planetesimals accreted into the original Earth, the large ones formed the Earth's central parts, and the small ones its outer parts [15, 24, 29]. U, Th, and other heat-producing elements relatively enriched in the crust of large planetesimals, and thus, they would relatively enrich in the contact positions (zones) between planetesimals during Earth's accretion as indicated in fig.2.

**The formation of ultra mafic and basaltic crusts:** in the early stage, the energy from the impacts between planetesimals, and of the part from the radioactive decay energy released by U, Th, K, and short life radioactive elements, reached its largest, and mostly located at the contact positions between planetesimals. These energies raised the temperature of the rocks at the contact zones between planetesimals to the highest, made

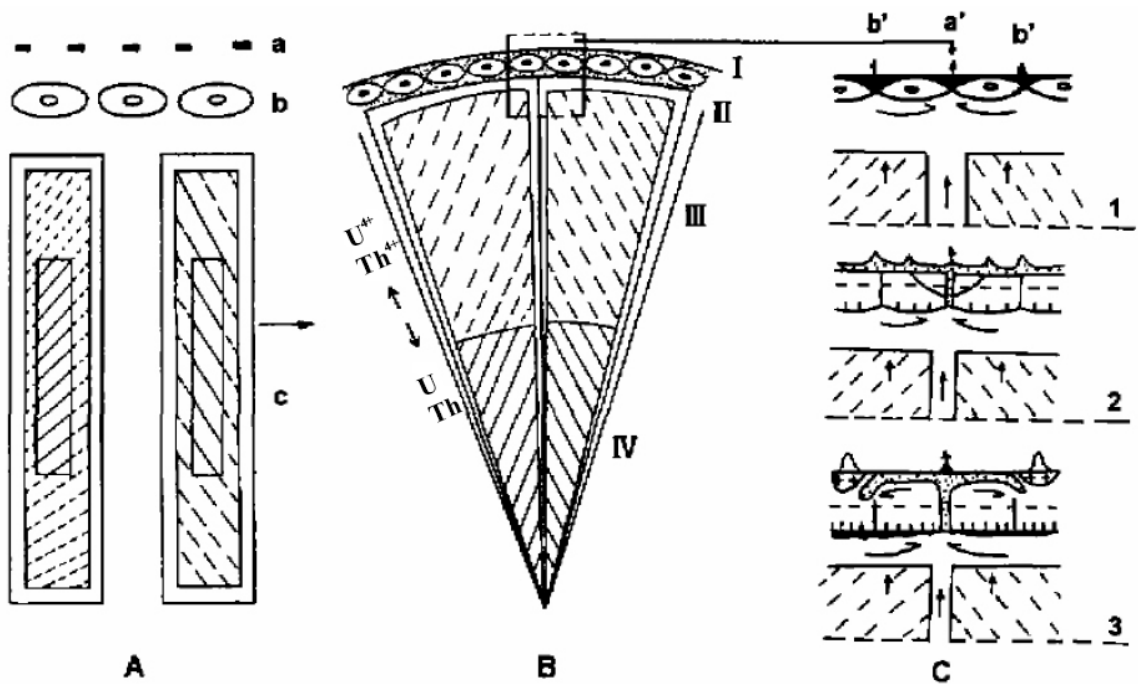

**Fig 2 The migration of U and Th and the origin and evolution of the enrichment zone (EZ) and its influence on the Earth's crust during and after the formation of Earth from planetesimals**
A: a- radius >10 km planetesimals; b- radius > 100 km planetesimals with original core and mantle; c- radius > 1000 km planetesimals with original core, mantle and crust.
B: I-original lithosphere; II-enrichment zone of U, Th and other heat-producing elements. III-original mantle; IV-original core ($U^{4+}$ and $Th^{4+}$ signify the high valence U and Th compounds and complexes that have a tendency to migrate from high to low temperature phases; U and Th signify the metal U and Th, and their stable low valence oxides and compounds)
C: 1- ultra mafic and basaltic magma eruption stage; 2- **granite magma eruption stage;** 3-original plate subduction stage (arrows signify the migration directions of U and Th, the dark areas indicate the ultra mafic rock layers) them melt at a rate of 100%, and formed superheat magmas (> 1680 $^{o}$C). These magmas moved up at a speed of several meters/second along with the weak positions in the contact zones between planetesimals, and erupted onto the surface. After reaching the surface, magmas would not have enough time to differentiate, but rapidly crystallized. Consequently, komatiites with special lish structures [31] and undifferentiated ultra-mafic primitive crusts formed from these magmas as shown in fig. 2C-1. Afterwards, the energy from the impacts between planetesimals and the decay energy from the short-life radioactive elements sharply decreased. Consequently, the temperature in the contact zones between planetesimals dropped and then the basaltic magma formed and erupted to the surface to form basaltic crusts.

**The formation of granitic crusts:** With the passage of time, the heat released from U, Th and other heat-producing elements would have raised the Earth's interior temperature. Volatile and water components would be gradually released from minerals in rocks and migrate up. As mentioned above, they would oxidize U and Th and form high valence compounds and complexes with small densities and low melting points. These high valence compounds and complexes would migrate up because of their low densities and low melting points. During the process of migration, the heat released from U, Th and other heat-producing elements would melt the granitic compositions with low melting points, and make them migrate up. This kind of migration mainly occurred along the structurally weak positions in the contact zones between the original planetesimals. When they reached the surface, the volatile and water components formed the atmosphere and oceans, and the meltable components and heat producing elements U, Th etc. formed tremendous granitic rocks. During this process, granitic components first reacted with, and then assimilated the original ultra mafic and basaltic crusts. Some of these older crusts survived and became inclusions in archaeozoic granitic rocks. For instance, there are about 20% dark components (mianthites) with high ultra mafic compositions in the archaeozoic granitic rocks located in Inner Mongolia and Hebei, China [17]. At the same time, the mafic components in these granitic rocks are also relatively high [17]. They are usually characterized with compositions of charnockites with a high content of hypersthene [17]. We also found that the zircons from several 3.8 Ga old granitic rocks are relatively high in mafic elements, such as Ni and Cr [19].

The surface distribution of the granitic magma passages inherited from the contact zones between the original planetesimals is inhomogeneous (a' and b' positions in fig.2 c). Consequently, the distribution of granitic rocks in the crust is also inhomogeneous. From this point, a pangea or a single continent composing of granitic rocks in the original crust is unnecessary. The shape of the contact zones between original planetesimals is mainly dot like, so this can explain why the structures of archaeozoic granitic rocks are mainly oval [32].

In this model, the formation of the granitic rocks does not need a magma ocean produced by a complete melting of the original crust and mantle. Therefore, the original crust and mantle heterogeneities [29, 33] can be explained based on this model. Namely this kind of heterogeneities might inherit from the composition heterogeneities of original planetesimals. Furthermore, the calculated results indicate that the granitic components from the differentiation of a melted upper mantle are not enough to form the thick old granitic crusts we have had [16]. According to this model, the granitic components in the crust come from the contact zones between the original planetesimals not only within the original crust and upper mantle, but also within the original lower mantle and core. Therefore, the meltable granitic components from the whole Earth will be enough to form the granitic continental crusts we live on.

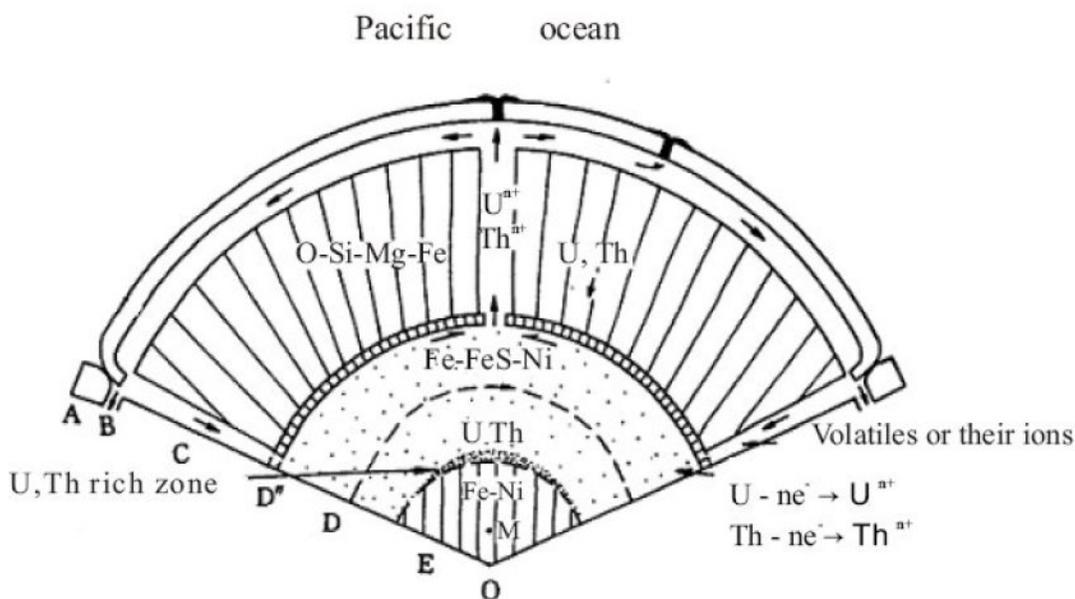

**Fig. 3 The distribution of U and Th in the outer core and its influence on the formation of deep mantle plumes and subducted lithosphere plates in the Pacific Ocean.**
A-lithosphere; B- U,Th enrichment zone or asthenosphere; C- mantle; D''- core-mantle boundary; D- outer core; E- inner core; O- center of Earth; M- center of geomagnetic field and thermal convection; the black points represent the relative concentrations of U and Th.

The U, Th rich zone may be located in (or close to) the bottom of the outer core under the hottest mantle beneath the Pacific Ocean. It may have been offering energy to create and maintain the Pacific superplume and been responsible for the opening of the Pacific Ocean. It may also cause the geomagnetic field center (M) to deviate from the Earth's center (O) to the Pacific Ocean direction. It is usually thought that the Earth's magnetic field is induced by the heat convection in the outer core. Therefore, we speculate that the magnetic center M overlaps with the heat center (or heat convection structure center) in the core that is closely related to the U, Th rich zone. If this U, Th rich zone migrates, then everything related to it may migrate correspondingly.  Details can be found in Bao (1999, Geological Review, Vol. 45, Sup. 82-92).

**The formation of a U and Th enrichment zone (EZ) and its relationship with plate tectonics:** when the meltable granitic components with a high content of heat-producing elements migrated upwards, the heat released from the heat-producing elements would melt granitic components in the original mantle and move up to the Earth's crust. Consequently, the remnant heat-resistant (refractory) components gradually built up ultra basaltic layers under the original lithosphere as indicated in 3 of fig.2C. The heat-resistant characteristic of these layers gives them the ability to resist the further migration of U and Th upwards. As a result, an EZ (3 in fig.2 C) with a high content of heat-producing elements was formed. This EZ might partially originate from the crusts of the big original planetesimals with a high content of heat-producing elements as indicated in fig. 2.  The content of heat-producing elements in the EZ was highest in the stage right after Earth's accretion. These heat-producing elements not only originated from the differentiated crusts of the big planetesimals, but also come from their greater abundance in the original planetesimal material, including long-life and short-life radioactive elements (their abundances were highest right after the formation of the solar system, and then declined with time). Adding the heat produced by the impacts between the original planetesimals during Earth's accretion, the temperature at the new born enrichment zone would reach its highest. This would produce a thin lithosphere due to the fact that a lot of rocks had been transformed into magma by the super-high temperature. Thus, in the early stage, volcanic activities dominated the Earth's surface. With the formation of the granitic crust above, a large amount of U, Th and K migrated from the EZ to the crust, which would lower the temperature of the EZ and thicken the lithosphere. At the same time, the heat-resistant ultra-basaltic rock layers in the bottom of lithosphere were thickening with the loss of the meltable granitic components from here. Consequently, volcanic activities were gradually limited to the former contact zone positions between the original planetesimals emerged on the surface as marked in 3 in fig.2C and fig. 3, and the plate tectonics began develop locally. Namely, small rigid plates began to form, horizontal movement appeared locally, and plate subductions occasionally occurred. When the lithosphere thickened and developed to a global scale, the heat energy released from U, Th and other heat-producing elements in the EZ would make a global plate tectonic system form and function. First, with the resistance of a thickening lithosphere with a heat-resistant ultra-basaltic rock layers in its bottom, the U and Th would gradually accumulate in the EZ. When their content reached a high level, they would heat up the rocks in the EZ close to melting. This makes the rock plastic or malleable [34]. Consequently, the horizontal movement of the overlying lithosphere plates becomes possible. Secondly, the heat released from U and Th also produced magma continually. The magma would move up along the oceanic ridges, produce new oceanic crusts there, and pushed the oceanic plates horizontally. At oceanic trench positions, the rigid oceanic plates with plenty of water and volatile components would subduct to the mantle and even core mantle boundary (CMB) [15,16]. Then these water and volatile components could react with U and Th in the interior (including the mantle and core) and make them migrate up to the EZ. The U and Th in the EZ drive the plate tectonic system and allow it to develop and evolve.

    The magma eruptions mentioned above might have been intermittent and catastrophic in the early stage, since only then the U, Th and other heat-producing elements in the EZ are rich enough to release heat to produce a tremendous amount of magma. When the accumulated magmas have enough power to break through the overlying rock layers, they erupt to the surface. These kinds of violent magma eruptions happened at least in the early stage as indicated in 1 and 2 in fig.2 C. Until the rigid lithosphere plate formed, the magma eruptions changed gradually from "intermittent and catastrophic" volcanic activities to "uniform" transfers in a plate tectonic system as shown in 3 of fig.2 C. The magma eruptions shown in the 1 of fig.2 C only happened in the early crustal formation stages. However, the magma eruptions in stages 2 and 3 in fig. 2 C would coexist for a long geological time. In the early stage, magma erupted mainly intermittently and catastrophically. However, after around 1.0 Ga, magma began to migrate to the surface gently in a plate tectonic system. The important imprint of which is the occurrence of the Wilson cycle [29]. After the phanerozoic era (0.6 Ga), magma transferring from the interior to the surface was gradually dominated by a gentle way in a plate tectonic system. Afterwards, life began to develop and evolve rapidly. The contact zones between the original > 1000 km planetesimals might have become the passages to

the EZ for U and Th from the core and the deep mantle. Some of them might be inherited to form today's super plumes. Currently, this kind of migration is still ongoing, as such magma transfer can be observed directly at locations like Hawaii. The U and Th, and heat materials from the core and deep mantle, together with those in the EZ form the main energy sources for the crust and plate movements.

## 4. THE UNIQUENESS OF EARTH AND ITS RELATIONSHIP WITH U AND Th

Compared with other terrestrial planets, Earth is the only planet to have developed oceans and life [34]. Furthermore, it also has the following unique features: an active global plate tectonic system characterized by many folded mountains produced by the horizontal movements of plates, unique globally distributed oceanic ridge and trench systems [34], completely differentiated granitic continental crusts and so on. These indicate that our planet has plenty of energy in its interior to drive its crustal movements and evolution.

With the exception of meteorite impact events, Mercury and Moon have been geologically inactive since 3.0~4.0 Ga. Similarly, almost half of the surface of Mars is occupied by 3.0 ~ 4.0 Ga old meteorite-impact craters, there may have been some new large volcanic activity in the recent 100 ~ 200 Ma in limited areas though. However, no phenomenon produced by horizontal movement (one of the indicators of plate tectonic movements) has been discovered [34]. Also, folded mountains have not been discovered on Venus. When compared with Earth, its surface is extremely flat [1]. It appears that there is no workable mechanism to develop a plate tectonic system in these planets.

According to the differentiation characteristics of the Sun's protoplanetary disk, from Mercury, Venus, and Earth to Mars, the compositions of their building materials changed regularly, namely U and Th increased gradually [22]; volatile components also increase gradually [1]. After becoming planets, from Mercury, Venus, and Earth to Mars, the ratio of iron meteorite components with a low U and Th [10] decreased gradually, but stone meteorite components with a relatively high U and Th, accompanied by a high content of volatile components, increased gradually. Consequently, the ratio of iron meteorite components is the highest in Mercury, and then decreases in order from Venus, Earth to Mars. Therefore, the U and Th in Mercury and Venus are relatively low, and their interior energy is relatively small. This may be one of the reasons why there have been no strong volcanic activity on Mercury since 3 ~ 4 Ga. It also is one of the reasons why the differentiation degree of Venus is much poorer than that of Earth, and no plate tectonics developed there.

However, the distance from the Sun is the same between Earth and the Moon, and much larger for Mars. Thus the U and Th content in the Moon and Mars should be similar or higher than that of Earth, but both Mars and the Moon have almost been dead since 4.0 Ga. The reason may be that there is not a workable mechanism in Mars and the Moon to use the energy released from U and Th inside their interiors. A lack of oxidative volatile components may be an obstacle for Mars and the Moon to form such a workable mechanism.

It is well known that Earth is rich in volatile components; one of the obvious indicators is the wide oceans on its surface. These do not exist on the Moon and Mars.

The major reason for a lack of oxidative volatile components is their small masses; the moon is only 1/81 the mass of Earth and Mars is only 1/10 the mass of Earth [3]. The small masses of the Moon and Mars mean that their gravity was not strong enough to prevent volatile components from escaping during formation or impact events. For instance, Venus is similar to Earth in mass and it has a dense atmosphere with an atmospheric pressure of 92 times of the Earth's, but it has no oxidative volatile components with 96% $CO_2$ [1].

In Mars or our Moon, U and Th have existed in low valence compounds or metals with large densities because of a lack of oxidative volatile components. Thus they would have difficulty migrating upwards, but instead sank down to the deep interior and the core as illustrated in fig. 2 B. In this case, an EZ under the lithosphere can not be formed. Consequently, there is no dynamic mechanism to drive the crustal horizontal movement and develop a plate tectonic system.

In the early stage, partial U and Th might have been sequestered into the core during the early core segregation event(s). However, the remnants of U and Th might still be distributed in the silicate part of a planet. Because there is not a plate tectonic system to release the interior heat, including the decay heat from U and Th, the planet core temperature would increase with time. This would lead to more silicate mantle melting and more metals with U and Th and their low valence oxides and compounds being separated from the mantle to the core. This would cause the planetary interior temperature to increase. In order to release the accumulating heat, huge super-plumes might have been developed. For instance, the mass of Mars is only 1/10 of the Earth, but it has the largest volcanic systems in the solar system. The large volcanic systems may be the result of developing super-plumes in Mars. Similarly, a lack of oxidative volatile compounds may be the major reason for Venus having not developed a plate tectonic system, since it also could not form an EZ with a high content of U and Th without the help of the oxidative volatile compounds.

Likewise, Earth lacks oxidative volatile components in its deep interior and core. In where, U and Th can only exist as stable low valence compounds, metal or alloys like $U_6Ni$ [9]. At the beginning, they might enter the Earth's center, and the heat released from them could help to keep the metal core liquid. Afterwards, they can not alloy with metal Fe, Ni in the centre because of their large atomic size and as a result, migrated outwards. After losing the U and Th, the inner core would decrease in temperature and crystallize into solid metals. Therefore, metal U and Th and their low valence compounds may still exist in the liquid outer core as indicated in fig.3. This liquid status of the outer core has been confirmed by the seismological data [3]. We think that the U and Th in the outer core can combine with the oxidative volatile components from the mantle, subducted slabs, or from the release of the crystallizing inner core, become moveable U, Th compounds and complexes, and migrate to the EZ through super plumes. They continually offer energy for the Earth's interior material and crust movements.

## 5. THE POSSIBLE INFLUENCE ON THE DEVELOPMENT AND EVOLUTION OF LIVING BEINGS

The original ocean might have formed four billion years ago (4.0 Ga) [35]. Life might have begun 3.5~3.8 Ga ago [36]. However, the life had been kept simple until 0.615~0.52 Ga (Cambrian period), when life began to develop and evolve tremendously. Since then, explosive booms have been happening in life development and evolution.

From here, it appears that there is a great unbalance in the development of life and its evolution throughout geological history: namely in its first 2.9-3.1 Ga, life existed only as lower order bacteria and stromatolites, etc. After that, in a short time of 0.6 Ga, life has undergone huge developments and evolution from lower to higher order plants, from lower (began from ~0.7 Ga) to higher order animals, and finally to human beings.

According to the theory of evolution, formalized by Charles Darwin, natural selection and adaptation (or survival of the fittest) is the cause for life evolution [37]. Namely, environmental variation would lead species to produce continuous variation, and these kinds of variations can lead to life evolution by accumulating through a long time.

However, from molecular biology, heredity is transmitted through DNA gene molecules. DNA replicates itself faithfully and gets passed on to the subsequent generations. From this point, DNA can not produce any evolution [37, 38].

In modern molecular biology, it is thought that life evolution depends on DNA mutations. Namely new species are created by the mutations of DNA, and the DNA mutation can not be formed from environmental variations [37, 38].

From the reproductive invariance of genes (DNA) or the functional utility acquired by organisms, the occurrences of mutations are blind and accidental, due to chance [from Jacques Monod in refs 36, 37]. Chance is the sole source of every innovation, of all creation in the biosphere.

We think that now that life development and evolution has happened on the Earth's crust, the "chance" described above should be controlled by the "necessity" of the crustal geological evolution.

Many experiments have confirmed that X-rays can produce DNA mutations [39, 40], and have broadly been used in genetics research. For instance, T. H. Morgan found that X-ray radiation from radium can induce heritable mutations in fruit flies [38]. Therefore, there is no doubt that X-rays can induce DNA mutations, and U and Th are the very natural sources of X-rays.

The DNA mutations produced by U and Th radiation can be found among workers who mined uranium ore. An investigation showed that 52~58% of the investigated uranium ore miners died of cancers, and their cancers are heritable. This kind of large mutations of DNA is produced by X-rays released from U [41].

Nerueher pointed out that the variation of U concentration in sea water may be one of the causes of species extinctions [4]. For instance, U can reach 2500 parts per million (ppm) and Th can reach 4500 ppm in some fish fossils [5,6,10]. Coals, the fossils of plant, can contain 20-80 ppm of U and 100 ppm of Th [5, 6, 10]. Their death may be caused by the radiation released by the high level of U and Th.

These facts indicate that radiation from U and Th can produce mal-mutations, even the death of living beings. On the other hand, granites, usually used as building materials, also contain U (~4ppm) and Th (8.5-17 ppm), but they are harmless to human and other living beings. Therefore, a low dose of X-rays from U and Th can not mutate the DNA of humans and other living beings. Therefore, we think that in geological history, radiation levels produced by U and Th between the two extreme cases above, namely in uranium ore, and in granites, should have occurred somewhere or sometime. These radiation levels can produce small DNA mutations, but they are not deadly.

According to Mendel's theory of hereditary particles, even this kind of mutation occurred in an individual, and this kind of mutation can be passed on to a population by mating. This kind of small mutation can build up a source by accumulating in a long geological time, and produce biodiversity and drive the evolution of life [37, 38].

As discussed before, U and Th migrate gradually from Earth's interior to the crust. However, the ways they migrate is different from time to time: in a considerably long time after Earth's formation, violent volcanic eruptions dominated the crust as indicated in 1 and 2 in fig. 2C. During these periods, the increase of U, Th and volatile components are catastrophic, and each volcanic eruption would increase the concentration of U and Th and other shorted-life radioactive elements to an unprecedented level. In this case, the radiation from these radioactive elements would produce large and deadly mutations or cause extinction of living beings, and have prevented or slowed down the development of life and evolution. From here it is understandable that although life began 3.5~3.8 Ma ago, it boomed only 0.6 Ga from now. A global plate tectonic system began to build up after 1.0 Ga as indicated by the occurance of Wilson's cycle [29], and dominated the Earth's crust after 0.6 Ga (Phanerozoic era). During these newer stages, large volcanic activities were limited to small high heat release areas. The migration methods of U and Th shifted gradually from violent volcanic eruptions to gentle transfers done by a plate tectonic system as discussed above. In other words, after 0.6 Ga, the increase of U and Th in the crust is gentle and limited, similar to today's rate. This kind of increase can not create a catastrophic impact on living beings. On the contrary, this kind of gentle and gradual increase of X-ray dose from U and Th may induce small DNA mutations. An accumulation of these mutations could become an important source to create biodiversity and support development and evolution in biosphere.

After the discovery of X-ray mutagenesis in living beings with varied doses of X-rays, H. J. Muller proposed that humans can improve themselves by X-rays to adapt the environment of extraterrestrial planets [40]. Namely, weak natural X-rays can be used to produce beneficial/useful mutations (warning: it is only a hypothesis, not to be tested on human beings). Consequently, we think the content variation of U and Th in the crust may be one of the major causes for the development and evolution of living beings.

## 6. CONCLUSIONS

(1) Under the lithosphere, there is an enrichment zone (EZ) of U, Th, K and other heat-producing elements. It offers energy to produce crust and drive its evolution. It also creates a plate tectonic system and makes it workable.
(2) In the parts with oxidative volatile components of the mantle, U and Th exist as high valence compounds and complexes, and gradually migrate to the EZ.
(3) In the interior without oxidative volatile components, including lower part of the lower mantle and the core, U and Th have existed as low valence oxides and compounds, and metals. They have sunken to the deep interior and are partially sequestered to the core, and are currently located in the outer core. However, at the core mantle boundary they could combine with remnant oxidative volatile components or ions to form movable high valence compounds and complexes to migrate to the EZ through upward plumes. These oxidative volatile components or ions may come from the remnants in the lower mantle, the subducted slabs and downward plumes in a plate tectonic system.

(4) Earth is the only terrestrial planet rich in $H_2O$, $O_2$, $N_2$ and other oxidative volatile components or ions. These oxidative volatile components have participated to and accelerated the migration of U and Th. They made the U and Th content of the EZ reach its highest level in the early stages of Earth, and then gradually lowered to the current level. Correspondingly, geological activities were dominated by violent volcanic eruptions in the Earth's early stages. Until 1.0~0.6 Ga, geological activities were changed to gentle plate tectonic movements. Since 0.6 Ga, plate tectonic activities have dominated the Earth. Therefore, the intensity of geological activities declined with geological time in the Earth, which is favorable to the development and evolution of life. In Mercury, Venus and Mars, U and Th would exist as low valence oxides and compounds, and metals, and mostly sink to their cores because of a lack of oxidative volatile components or ions. Consequently, an EZ can not have been formed in these planets. Therefore, there is no global crustal movement, and also no plate tectonic system developed in these planets. However, the heat in their cores can be released by strong volcanic activities through super plumes. It is also possible that the accumulating heat in their cores will melt or partially melt their mantle, and then create mantle convection, global crustal movements and plate tectonic systems in the future. Therefore, planets lacking oxidative volatile components or ions have a much slower evolution and are characterized by a rising trend with time in crustal movements, which is contrary to that of our planet.

(5) The strong volcanic activities in Earth's early stage added the U, Th and other radioactive elements to the crust rapidly. The radiation from these radioactive elements would destroy the DNA molecules of life beings, which has limited or stopped the development and evolution of life in the early stages. After the Phanerozoic era (0.6 Ga), the transfer of U and Th mainly is completed by gentle plate tectonic systems. Consequently the increase of U and Th in the crust has become gentle. A gentle increase in radiation levels from U and Th can not destroy life, but slightly mutate the DNA molecules of life beings. This may create biodiversity and drive the development and evolution.

**Notes:**

This is our paper that was originally published in ACTA PETROLOGICA ET MINERALOGICA, June, 1998, Vol. 17, No. 2, 160-172.

New development including experimental evidences in Earth and terrestrial planetary dynamics, please refer to:

1. Xuezhao Bao, 2006. Uranium solubility in terrestrial planetary cores: evidence from high pressure and temperature experiments. PhD thesis of the University of Western Ontario, pp174.

Or three chapters from it:

2. Bao, X., Secco, R.A., Gagnon, J.E., Fryer, B.J., 2006. Uranium partitioning between liquid iron and silicate melt at high pressures: implications for uranium solubility in planetary cores. At http://arxiv.org, arXiv:astro-ph/0606614, June 2006

3. Bao X., Secco, R.A., 2006. U solubility in Earth's core. At http://arxiv.org, arXiv: astroph/0606634, June 2006.
4. Xuezhao Bao, 2006. Terrestrial planetary dynamics: a view from U, Th geochemistry. At http://arxiv.org, arXiv:astro-ph/0606468, June 2006.

Similar views in the development and evolution of life can be found in:

Parnell, J. 2004. Plate tectonics, surface mineralogy, and the early evolution of life. International Journal of Astrobiology 3 (2) : 131–137.

**Acknowledgements**

I would like to thank Dr. RA Secco for offering me a postdoctoral position, which provoked my interest in translating this paper in my after-work time.

**References**

1. Chinese space science society, 1987. The dictionary of space sciences. Beijing: Science Press, 116-164.
2. Huo Defeng, Oyan Zi Yuan, 1974. Nucleus transit energy and the evolution of the Earth's matter. Beijing: Science Press. 1-91.
3. Wuhan College of geological sciences, 1979. Geochemistry, Beijing: Geological Publishing House.
4. Willians G. E. (ed.), 1986. Big cycle. Translated by Ma zhongjin et al. Beijing: Seismology Press. 141-142.
5. Liu Yingjun, 1984. Element geochemistry. Beijing: Science Press, 1-228.
6. Liu Yingjun, Chao Liming, 1987. An introduction to elemental geochemistry, Beijing: Geological Publishing House, 81-97.
7. Culbicciotti EHP, 1977. Uranium chemistry. Beijing: Atomic Energy Press, 1-238.
8. Slavinsky, MN, 1959. The physical and chemical properties of elements. Translated by Huan Zhangtian. Beijing: Metallurgy Press. 1-122.
9. Cotton, F.A., Wilkinson, G., 1980. Advanced inorganic chemistry (2). Translated by the Beijing normal university. Beijing: People's Education Press. 1-775.
10. Rogers JTW, Adams TAS, 1976. Handbook of uranium and thorium geochemistry. Translated by Xiao Xuejun. Beijing: Atomic Energy Press. 1-103.
11. Krass, AA, Nobayev, 1980. The geochemistry and origin of Zirconia in kimberlite. Translated by Chu Hebao. Geology-Geochemistry. 12, 41-46.
12. Zhang Weizhao, Wang Henian, Wang Manyun, 1987. Ligand chemistry and its application in geology. Beijing: Geology Press. 1-199.
13. Hu Zhonggeng, Chen Jiazhong, 1992. An introduction to chemistry for geology. Beijing: Higher Education Press, 1-141.
14. Claude-Jeanallegre, 1979. An introduction to geochemistry. Translated by Zhi Xiachen. Beijing: Geology Press. 70-90.
15. Huo Wei, Xie Hongsen, 1996. A discussion on the behavior of water in Earth's evolution. The Advance of Geosciences, (4): 350-355.
16. Wang Hongzhen, 1997. Speculations on earth's rhythms and continental dynamics, Earth Science Frontiers. 4(3-4): 1-11.
17. Bao Xuezhao, Lu Songnian Li Huiming, Li Huaikung, 1995. Minerageny of zircons from high-grade metamorphic rocks in inner Mongolia and Hebei. Acta Petrologica Et Mineralogica, 14( 3): 252-261.

18. Bao Xuezhao, 1995. Two kinds of composition variation treads of zircons and their significance in origin interpretation. Acta Mineralogica Sinica. 15(4): 404-410, or at http://arxiv.org, arXiv: 0707.3180, July 2007.
19. Bao Xuezhao, 1996. A study in the minerageny of zircons from the remnants of 3800 Ma Crust in China, Acta Mineralogica Sinica, 16(4): 410-415.
20. Bao Xuezhao, Li Huimin, Lu Songnian, 1997. Genetic mineralogy of zircon and its application to eclogite. Acta Petrologica Et Mineralogica, 16: supplement 90-96.
21. Bao Xuezhao, Lu Songnian, Li Huiming, 1998. A Raman spectroscopic study of zircons on micro-scale and its significance in explaining the origin of zircons. Scientia Geologica Sinica, 33(4): 473-480, or at http://arxiv.org, arXiv: 0707.3184, July 2007.
22. Wang Xiang, Pupin, JP, 1992. Element geochemistry of zircons and geologlcal implication for Argentera granite, france. Geological Review, 38 (3): 260-270.
23. Guo Shilun, Hu Ruiying, 1995. Uranium content distribution and migration in Chondrite, Geology Geochemistry. (4): 121-125.
24. Hou Wei, Ouyang Zhiyuan, 1995. A new idea on exploring the condensation of original protoplanetary cloud. Chinese Science Bulletin. 40(14): 1294-1297.
25. Tatsumi Y, Hamilton DL, Nesbitt RW, 1986. Chemical characteristics of fluid phase released from a subducted lithosphere and origin of arc magmas: evidence from high-pressure experiments and natural rocks. Journal of Volcanology and Geothermal Research. 29 (1-4): 293-309.
26. Ляфхович, BB, 1988. Minerals are the indicators of lithosphere evolution. Translated by Li Zengtian. Geology Geochemistry. (3): 16-24.
27. Ringwood, AE, Kesson, SE, Hibberson W, Ware N, 1992. Origin of kimberlites and related magmas. Earth and Planetary Science Letters. 113 (4): 521-538
28. Wang Shijie, Ouyang Ziyuan, Zhang Fuqin, 1995. A discussion on the formation and initial state of the Earth--a two stage non-homogeneous accretion model. Geology Geochemistry. (5): 68-73.
29. Ouyang Ziyuan, Zhang Fuqin, 1995. A theoretical model on Earth's composition heterogeneity—a discussion on some problems in geology and geochemistry. Geology geochemistry. (5): 68-73.
30. Thones, JA, 1995. The origin of Earth. Translated by Cai Huamei. Geology Geochemistry. (5): 16-21, 23.
31. Wu changhua, 1995. An introduction to the book "Archean Crustal Evolution" written by Condie et al. Progress in Precambrian Research. (2): 30-40.
32. Zhang Qiusheng, 1984. Geology and Metallogeny of the Early Precambrian in China. Changchun: Jilin People's Publishing House, 1984. 1-400.
33. Xiao Zhifeng, Ouyang Ziyuan, Lin Wenzhu, 1995. The restriction of Earth's original non-homogeneity to the distribution of super ore deposits. Geology Geochemistry. (5): 75-80.
34. Hambin WK, 1980. The earth's dynamic systems. Translated by Yin Weihan et al. Beijing: Geological Publishing House. 1-408.
35. Liu Chaohui, Ouyang Ziyuan, 1995. The periodicity and its formation mechanism of the Earth's environmental evolution. Geology Geochemistry. (5): 81-89;
36. Fan Shiqing, 1982. Earth and ocean. Beijing: Science Press. 74-153
37. Chen Rongxia, 1996. The steps in evolution. Beijing: China social sciences press. 1-164.
38. Zhao Gongmin, 1996. The concept of heredity. Beijing: China Social Sciences Press. 1-170.
39. Yang Jianye, Zhu Xianmin. 1992. Dictionary of the Winners of Nobel Prize. Changsha: Hunan Scientific & Technical Publishers, 1-320.

40. Li Chengyuan. 1981. An introduction to the winners of the Nobel Prize in Physiology or Medicine. Beijing: Popular Science Press.
41. Cheng Weinan, 1992. Professional cancers and their prevention and control. Beijing: China Medicine and Technology Publishing House.